\documentclass[preprint,showpacs,aps,preprintnumbers,amsmath,amssymb]{revtex4-1}

\usepackage{graphicx,epsfig}
\usepackage{dcolumn}
\usepackage{bm}
\newcommand{\bpmat}{\begin{pmatrix}}
\newcommand{\epmat}{\end{pmatrix}}
\newcommand{\bwt}{\begin{widetext}}
\newcommand{\ewt}{\end{widetext}}
\newcommand{\beq}{\begin{equation}}
\newcommand{\eeq}{\end{equation}}
\newcommand{\beqa}{\begin{eqnarray}}
\newcommand{\eeqa}{\end{eqnarray}}
\newcommand{\bdm}{\begin{displaymath}}
\newcommand{\edm}{\end{displaymath}}
\newcommand{\lslash}[1]{#1\llap/}

\newcommand{\calE}{{\cal E}}
\newcommand{\Eq}[1]{Eq.\ (\ref{#1})}
\newcommand{\Eqs}[2]{Eqs.\ (\ref{#1}) and (\ref{#2})}
\newcommand{\Eqss}[3]{Eqs.\ (\ref{#1}), (\ref{#2}) and (\ref{#3})}
\newcommand{\Eqsss}[4]{Eqs.\ (\ref{#1}), (\ref{#2}), (\ref{#3}) and (\ref{#4})}
\newcommand{\Eqsto}[2]{Eqs.\ (\ref{#1})-(\ref{#2})}

\newcommand{\Section}[1]{Section\ \ref{#1}}

\begin{document}

\title{Neutrino propagation in an electron background with an inhomogeneous
  magnetic field}
\author{Jos\'e F. Nieves$^{a}$}
\email{nieves@ltp.uprrp.edu}
\author{Sarira Sahu$^{b,c}$ }
\email{sarira@nucleares.unam.mx}
\affiliation{$^{a}$Laboratory of Theoretical Physics,
  Department of Physics, University of Puerto Rico 
  R\'{\i}o Piedras, Puerto Rico 00936}
\affiliation{$^{b}$Instituto de Ciencias Nucleares, Universidad Nacional Aut\'onoma de M\'exico, 
Circuito Exterior, C.U., A. Postal 70-543, 04510 Mexico DF, Mexico}
\affiliation{$^{c}$Astrophysical Big Bang Laboratory, RIKEN, Hirosawa, Wako, Saitama 351-0198, Japan}

\begin{abstract}
 We study the electromagnetic coupling of a neutrino that propagates
  in a two-stream electron background medium. Specifically, 
  we calculate the electromagnetic vertex function for a medium that
  consists of a \emph{normal} electron background plus another
  electron \emph{stream} background that is moving with a velocity
  four-vector $v^\mu$ relative to the normal background.
  The results can be used as the basis for studying
  the neutrino electromagnetic properties and various processes in such
  a medium. As an application, we calculate the neutrino dispersion relation
  in the presence of an external magnetic field ($\vec B$),
  focused in the case in which $B$ is inhomogeneous,
  keeping only the terms of the lowest order in $1/m^2_W$ and linear
  in the $B$ and its gradient.
  We show that the dispersion relation contains additional anisotropic terms
  involving the derivatives of $\vec B$,
  such as the gradient of $\hat k\cdot(\vec v\times\vec B)$,
  which involve the stream background velocity, and a term of the form
  $\hat k\cdot(\nabla\times \vec B)$ that can be present
  in the absence of the stream background,
  in addition to a term of the form $\hat k\cdot\vec v$ and
  the well known term $\hat k\cdot\vec B$ that arises in the constant
  $\vec B$ case.
  The derivative-dependent terms are even under a $CP$
  transformation. As a result, in contrast to the latter two just mentioned,
  they depend on the sum of the particle and antiparticle densities
  and therefore can be non-zero in a $CP$-symmetric medium in which
  the particle and antiparticle densities are equal.
\end{abstract}

\maketitle

\section{Introduction and Summary}
\label{sec:introduction}

The subject matter of this paper is the propagation of neutrinos in a medium
in the presence of an external electromagnetic field. There are various
problems of interest associated with this subject that have been well studied
in previous works. In most previous studies the interest has been
on a medium consisting of a thermal background of various particle species,
which can be taken to be at rest, in the presence of an external magnetic
field in the same frame, which is assumed to be homogeneous. In those studies
typically the focus is on the dispersion relation of a neutrino
that propagates in such environments. The assumptions underlying the previous
works do not allow us to consider situations in which the
thermal backgrounds of the different particle species
move with some relative velocity relative to each other,
and/or situations in which the external field is not homogeneous.

There are several reasons why considering the more general
situations just mentioned above are of interest. For example,
the propagation of photons in two-stream plasma systems
is a well studied subject in the context of plasma physics,
particularly with regard to the so-called two-stream instabilities
\cite{Shaisultanov:2011hc,Yalinewich:2010,Sironi:2015oza},
many aspects of which have been studied
both analytically and numerically\cite{Drake:2003,Boris:1970,McMillan:2006a,McMillan:2007b,Goldman:2008}.
In recent works, similar studies have been carried out for
\emph{magnetized} two-stream plasma systems \cite{Che:2009yh,Soto:2010,Oraevsky:2003cf}.
In these works the focus is typically the dispersion relation
of the photon when it propagates in the environment that is being
considered. The case of propagation through inhomogeneous plasmas has
also been studied \cite{Fitzpatrick2015}.
Several authors have studied the propagation of neutrinos in moving
media in the presence of an external electromagnetic
field\cite{Giunti:2014ixa,Nunokawa:1997dp,Bergmann:1999rz}. Also 
the effects of moving and polarized matter
on neutrino spin/magnetic moment oscillations and $\nu_L \rightarrow
\nu_R$ conversions are considered \cite{Lobanov:2001ar,Grigoriev:2002zr,Studenikin:2004bu,Arbuzova:2009uj}. In ref. \cite{Shaisultanov:2011hc}, the growth
rates for different instabilities of the relativistic ion beams
propagating through a hot electron background are studied analytically
and checked 
with numerical simulations.
This configuration can be  of relevance to study 
the relativistic, collisionless
shock structures in astrophysical scenarios
where oppositely directed particle beams (protons)
pass through an isotropic electron
gas\cite{Yalinewich:2010,Lominade:1979}. 

From a fundamental and conceptual point of view
the problem we want to consider is the counterpart for neutrinos.
The problem of the propagation of neutrinos in magnetized media
is relevant in several physical contexts, such as pulsars \cite{Kusenko:1996sr},
supernovas \cite{Sahu:1998jh,Duan:2004nc,Gvozdev:2005}
and gamma-ray bursts \cite{Sahu:2009ds,Sahu:2009iy},
where the magnetic fields are believed to have important implications.
Also the effects of stream neutrino background have been suggested as a
mechanism of large scale magnetic field generation in the hot plasma of the Early Universe\cite{Semikoz:2003qt}.
In those contexts, the effects of stream backgrounds and/or inhomogeneous
fields can be of practical interest.

In a recent work \cite{Nieves:2017rvx} we initiated
the study of the propagation of neutrinos in medium along these lines,
calculating the self-energy and dispersion relation of a neutrino
that propagates in a magnetized two-stream background medium.
Specifically, we considered a medium composed of an electron background,
which can be taken to be at rest, and a second electron background
that moves with a velocity four-vector $v^\mu$ relative to the first. We refer
to them as the \emph{normal} and \emph{stream} backgrounds, respectively.
In addition we assumed that, in the rest frame of the normal background,
there is a magnetic field ($\vec B$) that is homogeneous.
The calculation was based on the local limit of the weak interactions,
and therefore restricted to the leading $O(1/m^2_W)$ terms,
and on the application of the Schwinger propagator method,
adapted to the two-stream background, but keeping only up to the linear
terms in $\vec B$.

The main results obtained in ref. \cite{Nieves:2017rvx} are summarized as
follows. For a neutrino that propagates in a two-electron background
and a constant magnetic field, as described above,
the dispersion relation acquires an anisotropic contribution of the
form $\hat k\cdot\vec v$ (where {$\hat k$} is the unit vector
  along the incoming neutrino momentum $\vec k$), in addition to the well known term
$\hat k\cdot\vec B$\cite{DOlivo:1989ued,Erdas:1998uu,Kuznetsov:2005tq,Erdas:2009zh}.
As discussed and explained in ref. \cite{Nieves:2017rvx}, a term of the form
$\hat k\cdot(\vec v\times\vec B)$ does not appear in the
constant $\vec B$ case. The physical reason behind this
result is that such a term is odd under time-reversal and there
is no source of time-reversal breaking effects in the context
of our calculations. However it was noted that terms
of similar form, but involving the derivative of the electromagnetic
field, are even under time-reversal and therefore could be present
in the case that the electromagnetic field is not homogeneous.

As a continuation of the above work, here we calculate the electromagnetic
vertex function of a neutrino that propagates in a two-stream electron
background. This complements and extends our previous work in at least two ways.
On one hand, the knowledge of the vertex function allows
us to determine the neutrino electromagnetic properties and to
calculate the rate for various processes involving
neutrinos in such media, in analogy with the study of electromagnetic
neutrino processes in ordinary media \cite{Raffelt:1996wa}.

On the other hand, which we pursue here, by considering the effective
neutrino interaction with an external electromagnetic field,
the result for the vertex function is
used to determine the self-energy and dispersion relation of a neutrino
that propagates in the two-stream electron medium with an inhomogeneous
magnetic field. The self-energy and dispersion relation for the case
in which there is only one electron background can be obtained as a special
case, whether it is moving or at rest relative to the external magnetic field.
This work complements our previous calculation of the dispersion
relation based on the Schwinger propagator method, which is restricted
to a uniform magnetic field. The dispersion relation obtained
can be used as the basis for studying the effects of inhomogeneous
fields on the neutrino oscillations in several environments
such as pulsars, supernovas, and gamma-ray bursts that have
been considered in the literature cited, as well
as several related application contexts where
the inhomogeneity of the magnetic fields may have a prominent
role such as transition radiation induced by a magnetic
field \cite{Ioannisian:2017mqy}, neutrino-induced plasma instabilities
in supernova \cite{Yamamoto:2015gzz}, neutrino driven magnetic field
instability in a compact star\cite{Dvornikov:2013bca} and the effects of asymmetric
neutrino propagation in proto-neutrons stars \cite{Maruyama:2013rt}.

It is appropriate to mention here that the calculation of the
neutrino electromagnetic vertex function
in the two-stream electron background
is based on the local limit of the weak interactions, i.e.,
it is limited to the $O(1/m^2_W)$ contributions. Moreover,
in the application to the calculation of neutrino
self-energy and dispersion relation mentioned above we retain only
the terms that are at most linear in the derivatives of the field.

The results of the calculations confirm that in the case of an
inhomogeneous field the dispersion relation acquires
additional anisotropic terms that involve the derivatives of
the magnetic field. In particular, a term of the form
$\hat k\cdot(\nabla\times \vec B)$, which is independent of the stream
background velocity, can be present even in the absence of the stream
background. Other terms, such as the gradient of
$\hat k\cdot(\vec v\times\vec B)$ already mentioned above,
depend on the stream background velocity, but they can be
present even in the case in which  $\nabla\times \vec B = 0$.
Moreover, all these additional terms are even under the
$CP$ transformation and as a result they are proportional to the sum
of the particle and antiparticle densities. This is in
contrast with the $\hat k\cdot\vec v$ and $\hat k\cdot\vec B$ terms, which
are even under $CP$ and depend on the difference
of the particle and antiparticle densities. In situations
where the medium is $CP$-symmetric and
the particle and antiparticle densities are equal,
the $O(1/m^2_W)$ the contributions from the $\hat k\cdot\vec v$
and $\hat k\cdot\vec B$ terms vanish, and the contributions from the terms
involving the derivatives of the magnetic field could gain more
importance, depending on the degree of inhomogeneity of the magnetic field.

It is worth mentioning that, in order to calculate properly
the stream contribution to the vertex function, and more specifically
the vertex function's zero photon momentum limit (which is related to
the neutrino index of refraction), the screening effects
of the electron background must be taken into account.
The technical reason is that the electric
form factors (those that couple to the electric components of
the electromagnetic field $\sim \hat k\cdot(\vec v\times\vec B)$)
diverge in the zero photon-momentum limit,
unless the screening effects are taken into account.
In the case in which there is only one background ($\vec v = 0$),
or the magnetic field is uniform, only the magnetic form factor
couplings enter in the effective interaction
with the electromagnetic field, for which the screening corrections
are not relevant. An important ingredient of the present work
is the proper inclusion of the background screening effects that are present
in the kind of medium that we are envisaging,
in the calculation of the neutrino index of refraction.

In \Section{subsec:notation} we summarize some of the notations and
conventions that are used throughout.
The 1-loop formulas for the vertex function
are given in \Section{subsec:1loopvertex}, generalizing
the formulas given in ref. \cite{DOlivo:1989ued}, adapted to the present notation
and context. As already mentioned, they are
based on the local limit of the weak interactions, i.e.,
they are restricted to the $O(1/m^2_W)$ contributions.
The vertex function is expressed in terms of a set of
form factors that are given as integrals over the distribution functions
of the background electrons. Since the calculation of
the self-energy and dispersion relation in a non-homogeneous external field
involves evaluating the vertex function in the \emph{static limit}
appropriately, specially in the context of the two-stream system,
in \Section{subsec:staticlimit} we define precisely this limiting operation.
There we also summarize the static limit
value of the integrals involved in such formulas, which are relevant
in the calculation of the self-energy and dispersion relation.
Some of the calculation details regarding those formulas are provided in
Appendices
\ref{sec:ABCevalstaticlimit}, \ref{sec:Aprimeevalstaticlimit} and \ref{sec:CAprime0classical}.
The actual calculation of the self-energy in the presence of an external
inhomogeneous field is carried out in \Section{sec:selfenergy},
retaining the terms that are at most linear in the derivatives of the field,
and paying attention to the treatment
and incorporation of the screening effects of the electron background.
There we first enumerate the possible terms that may appear in the
$B$-dependent part of the self-energy under the specified conditions,
and write down its generic form in terms of a set of structure tensors
with corresponding coefficients to be determined. The coefficients are
then identified by considering the transition amplitude
in the presence of an external field, using the results of the
one-loop expression for the neutrino electromagnetic vertex function.
The need to include the screening effects for properly determining the
self-energy in the two-stream background case is explained there in more
detail. The corresponding dispersion relations are obtained and discussed
in \Section{sec:dispersionrelation}, focusing on some of the features
that illustrate the salient implications of the results for the
self-energy. In \Section{sec:conclusions} we review our work
and summarize the main results.

\section{The vertex function}
\label{sec:vertexfunction}

\subsection{Notation and conventions}
\label{subsec:notation}

We borrow some of the notation from ref. \cite{Nieves:2017rvx}, which we briefly
summarize here for convenience as follows. We use the symbols
$e$ and $e^\prime$ to refer to the electrons in the
normal and stream backgrounds, respectively,
while the symbol $f$ stands for either $e$ or $e^\prime$.
Denoting by $u^\mu_f$ the velocity four-vector of each background,
the convention stated above means that the velocity four vector
of the normal background is set to
\beq
\label{defue}
u^\mu_e = u^\mu\,,
\eeq
where, as usual,
\beq
\label{defu}
u^\mu \equiv (1,\vec 0)\,,
\eeq
while for the stream
\beq
\label{defueprime}
u^\mu_{e^\prime} = v^\mu\,,
\eeq
with
\beq
\label{defv}
v^\mu = (v^0,\vec v)\,.
\eeq

The relevant diagrams for the calculation of the electron background
contributions to the neutrino electromagnetic vertex are shown in
figure \ref{fig1}.
For the calculation we will need the following neutral current couplings
\beq
L_Z = - g_Z Z^\mu \left[
\bar e\gamma_\mu(X_e + Y_e\gamma_5)e +
\sum_\ell \bar\nu_{L\ell}\gamma_\mu\nu_{L\ell}\right] \,,
\eeq
where
\beqa
g_Z & = & g/(2\cos\theta_W) \,,\nonumber\\
X_e & = & -\frac{1}{2} + 2\sin^2\theta_W\,,\nonumber\\
Y_e & = & \frac{1}{2}\,.
\eeqa
We denote the momentum vectors of the incoming and outgoing neutrino by
$k,k^\prime$, respectively, and
\beq
\label{defq}
q = k^\prime - k\,,
\eeq
denotes the momentum vector of the incoming photon [
This convention is the opposite to the convention adopted
in Ref.~ \cite{DOlivo:1989ued} in which $q$ denotes the momentum of the outgoing photon.
This difference will is reflected in the sign of the $P$ term
in \Eq{defTTLP}]

The form factors of each background contribution are functions
of the scalar variables
\beqa
\Omega_f & = & q\cdot u_f\,,\nonumber\\
Q_f & = & \sqrt{\Omega^2_f - q^2} \,.
\eeqa
Physically, $\Omega_f$ represents the energy of the photon in the rest
frame of the normal background, while $Q_f$ is the magnitude of the 3-momentum
vector in the same frame, which we denote by $\vec Q_f$.

\subsection{1-loop formulas}
\label{subsec:1loopvertex}

\begin{figure}
{\centering
\resizebox*{0.35\textwidth}{0.35\textheight}
{\includegraphics{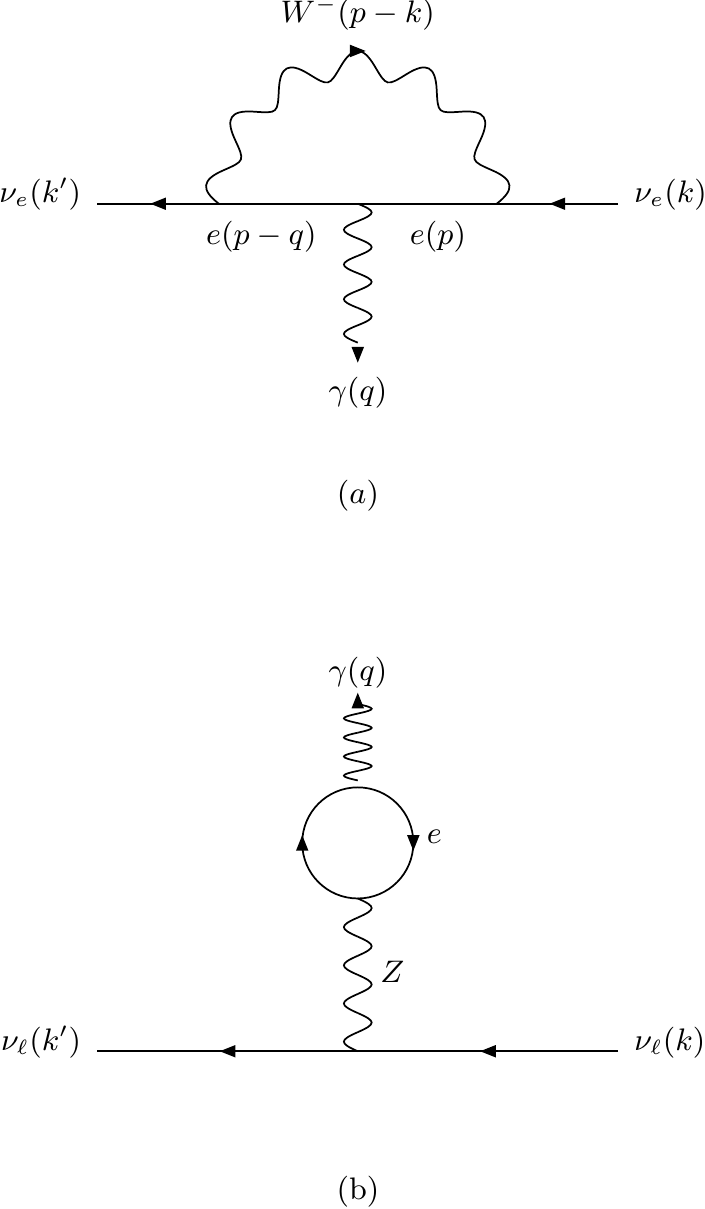}}
\par}
\caption{
The diagrams that contribute to the neutrino electromagnetic vertex
  in an electron background to the lowest order for a given neutrino
  flavor $\nu_\ell \; (\ell = e,\mu,\tau)$. Diagram (a) contributes
  only to the $\nu_e$ vertex function, while Diagram (b) contributes
  for the three neutrino flavors.
  \label{fig1}
}
\end{figure}

As already mentioned, the relevant diagrams for the calculation
of the electron background contributions to the neutrino
electromagnetic vertex function are shown in figure \ref{fig1}.
We denote by $\Gamma^{(W,Z)}_{f\mu}$ the contribution from
diagrams (a) and (b), respectively, and write the total
vertex function as
\beq
\label{defGammaWZ}
\Gamma_\mu = \Gamma_{e\mu} + \Gamma_{e^\prime\mu}\,.
\eeq
where
\beqa
\Gamma_{f\mu} = \left\{
\begin{array}{ll}
  \Gamma^{(W)}_{f\mu} + \Gamma^{(Z)}_{f\mu} & (\mbox{for $\nu_e$})\\[12pt]
  \Gamma^{(Z)}_{f\mu} & (\mbox{for $\nu_{\mu,\tau}$})
\end{array}\right.
\eeqa
%
%

We now rely on the results obtained in ref. \cite{DOlivo:1989ued}, adapted
for our present purposes. The results of the one-loop calculation of
$\Gamma^{(W,Z)}_{f\mu}$ is that $\Gamma_{f\mu}$ can be written in the form
\beq
\label{defGamma}
\Gamma_{f\mu} = T_{f\mu\nu}\gamma^\nu L\,,
\eeq
%
%
%
%
where the tensors $T_{f\mu\nu}$ do not contain any gamma matrices
and have the decomposition
\beq
\label{defTTLP}
T_{f\mu\nu} = T_{fT} R_{f\mu\nu} + T_{fL} Q_{f\mu\nu} - T_{fP} P_{f\mu\nu} \,,
\eeq
with $T_{eT,L,P}$ and $T_{e^\prime T,L,P}$ being scalar functions
of $\Omega_e,Q_e$ and $\Omega_{e^\prime},Q_{e^\prime}$, respectively. In
writing the last term in \Eq{defTTLP} we have taken into account the fact
that the  definition of $q$ here [\Eq{defq}] is the opposite to that in
ref.~\cite{DOlivo:1989ued}, as already mentioned in \Section{subsec:notation}.
%
%
%
%
The basis tensors $R_{f\mu\nu},Q_{f\mu\nu},P_{f\mu\nu}$
that appear in \Eq{defTTLP} are defined by
\beqa
\label{defRQP}
R_{f\mu\nu} & = & \tilde g_{\mu\nu} - Q_{f\mu\nu}\nonumber\\
Q_{f\mu\nu} & = & \frac{\tilde u_{f\mu}\tilde u_{f\nu}}{\tilde u^2_f}\nonumber\\
P_{f\mu\nu} & = & \frac{i}{Q_f}\epsilon_{\mu\nu\alpha\beta}
q^\alpha u^\beta_f\,,
\eeqa
where
\beq
\tilde g_{\mu\nu} = g_{\mu\nu} - \frac{q_\mu q_\nu}{q^2} \,,
\eeq
and
\beq
\tilde u_{f\mu} = \tilde g_{\mu\nu}u^\nu_f\,.
\eeq
The tensors $R_{f\mu\nu},Q_{f\mu\nu},P_{f\mu\nu}$ satisfy the relations
\beqa
R_{f\mu\nu}R^{\mu\nu}_f & = & -P_{f\mu\nu}P^{\mu\nu}_f = 2\,,\nonumber\\
Q_{f\mu\nu}Q^{\mu\nu}_f & = & 1\,,
\eeqa
while the contractions of anyone of them with the others vanish.

The functions $T_{fT,L,P}$ that appear in \Eq{defTTLP} are given by
\beq
\label{Ttotalforeachnu}
T_{fT,L,P} = \left\{
\begin{array}{ll}
  T^{(W)}_{fT,L,P} + T^{(Z)}_{fT,L,P} & (\mbox{$\nu_e$})\\[12pt]
  T^{(Z)}_{fT,L,P} & (\mbox{$\nu_{\mu,\tau}$})
\end{array}\right.
\eeq
where
\beqa
\label{TTLPformulas}
T^{(Z)}_{fT} & = & \frac{2eg^2_Z}{m^2_Z} X_e A^\prime_f\,,\nonumber\\
T^{(Z)}_{fL} & = & \frac{4eg^2_Z}{m^2_Z} X_e
\frac{B_f}{\tilde u^2_f}\,,\nonumber\\
& = & -\frac{4eg^2_Z}{m^2_Z} X_e \frac{q^2}{Q^2_f}B_f\,,\nonumber\\
T^{(Z)}_{fP} & = & -\frac{4eg^2_Z}{m^2_Z} Y_e Q_f C_f\,,
%
\eeqa
with
%
%
%
\beqa
\label{ABC}
A^\prime_f(\Omega_f, Q_f) & \equiv & A_f(\Omega_f, Q_f) -
\frac{B_f(\Omega_f, Q_f)}{\tilde u^2_f}\,,\nonumber\\
A_f(\Omega_f, Q_f) & = &
\int\frac{d^3p}{(2\pi)^3 2E}(f_{f} + f_{\bar f})\nonumber\\
&&\times \left[\frac{2m^2 - 2p\cdot q}{q^2 + 2p\cdot q} + (q\rightarrow -q)\right]
\,,\nonumber\\
B_f(\Omega_f, Q_f) &=&\int\frac{d^3p}{(2\pi)^3 2E}(f_{f} + f_{\bar f})\nonumber\\
&&\times
\left[\frac{2(p\cdot u_f)^2 + 2(p\cdot u_f)(q\cdot u) - p\cdot q}
{q^2 + 2p\cdot q} +\right. \nonumber\\
&&\left. (q\rightarrow -q)\right]\,,\nonumber\\
C_f(\Omega_f, Q_f) & = & \int\frac{d^3p}{(2\pi)^3 2E}(f_{f} - f_{\bar f}) \nonumber\\
&&\times
\frac{p\cdot\tilde u_f}{\tilde u^2_f}
\left[\frac{1}{q^2 + 2p\cdot q} + (q\rightarrow -q)\right]\,.
\eeqa
In these formulas, $m$ is the electron mass,
\beq
p^\mu = (E,\vec p), \qquad E = \sqrt{\vec p^{\,2} + m^2}\,,
\eeq
and $f_{f,\bar f}$ are the electron and positron thermal distribution functions
\beq
f_{f,\bar f} = \frac{1}{e^{\beta_f(p\cdot u_f \mp \mu_f)} + 1}\,,
\eeq
where $1/\beta_{e,e^\prime}$ and $\mu_{e,e^\prime}$ are the temperature and
chemical potential of the normal and stream background electrons,
respectively. The corresponding formulas for the
functions $T^{(W)}_{fT,L,P}$ corresponding to diagram (a) in
 figure \ref{fig1} are obtained from \Eq{TTLPformulas} by making the replacements
\beq
\label{WZreplacement}
\frac{g^2_Z}{m^2_Z} \rightarrow \frac{g^2}{2m^2_W}\,,\qquad
X_e \rightarrow \frac{1}{2}\,,\qquad
Y_e \rightarrow -\frac{1}{2}\,.
\eeq
From \Eq{Ttotalforeachnu},
\beqa
\label{Ttotalforeachnuexplicit}
T_{fT} & = & \frac{eg^2}{2m^2_W}A^\prime_f \times \left\{
\begin{array}{ll}
  1 + X_e & (\mbox{$\nu_e$})\\[12pt]
  X_e & (\mbox{$\nu_{\mu,\tau}$})
\end{array}\right.\nonumber\\[12pt]
T_{fL} & = & \frac{eg^2}{m^2_W}B_f \times \left\{
\begin{array}{ll}
  1 + X_e & (\mbox{$\nu_e$})\\[12pt]
  X_e & (\mbox{$\nu_{\mu,\tau}$})
\end{array}\right.\nonumber\\[12pt]
T_{fP} & = & \frac{eg^2}{m^2_W}C_f \times \left\{
\begin{array}{ll}
  1 - Y_e & (\mbox{$\nu_e$})\\[12pt]
  -Y_e & (\mbox{$\nu_{\mu,\tau}$})
\end{array}\right.
\eeqa

\subsection{Static limit}
\label{subsec:staticlimit}

As we have mentioned, the results for the electromagnetic
vertex function will be used as the starting point to determine
the self-energy and dispersion relation in an external field,
which involves evaluating the vertex function in \emph{static limit}. 
It is appropriate to state precisely what we mean by this limit,
specially in the context of our calculation that includes the effects
of the stream background and possibly a non-homogeneous external field.

Let us look first at the case considered in ref.~\cite{DOlivo:1989ued}, namely the normal
electron background contribution to the neutrino index of refraction
in the presence of an external constant $B$ field, that is a field that
is constant in time and homogeneous in space. This requires the evaluation
of the normal background contribution to the vertex function
in the \emph{zero momentum limit}, which operationally is implemented
by first setting
\begin{equation}
\Omega_e = 0\,,
\end{equation}
maintaining $Q_e$ fixed, and then taking the limit
\beq
Q_e \rightarrow 0\,.
\eeq
We indicate this two-step process by the notation
\beq
\label{staticlimit}
(\Omega_e = 0, Q_e \rightarrow 0)\,.
\eeq
The idealization involved here is that the $\nu\nu$
transition amplitude is being calculated over a region that
is microscopically large, but macroscopically sufficiently small
such that the external field is constant over the region.
In this situation, the terms in the $\nu\nu$ transition amplitude
that contain factors of $q$ multiplying the external field
do not contribute.

In the present work we consider the possibility that the external
field is not necessarily homogeneous. This is taken into account
by interpreting each factor of $q_\mu$ multiplying
the external field in the $\nu\nu$ amplitude as a derivative
\beq
\label{qasderivative}
q_\mu \rightarrow i\partial_\mu\,,
\eeq
acting on the external field.

In addition, in the presence of the stream, the limit $Q_e\rightarrow 0$
is complicated by the fact that the stream contributions to the neutrino
electromagnetic vertex function depend on the variables
\beqa
\label{OmegaQprime}
\Omega_{e^\prime} & \equiv & q\cdot u_{e^\prime} =
\Omega_e u^0_{e^\prime} - \vec Q_e\cdot \vec u_{e^\prime}\,,\nonumber\\
Q_{e^\prime} & \equiv & \sqrt{\Omega^2_{e^\prime} - q^2} \,,
\eeqa
where $\Omega_{e^{\prime}}$ represents the energy of the photon in the rest
frame of the stream background while $Q_{e^\prime}$
is the magnitude of the 3-momentum vector in that frame.
%
%
%
For $\Omega_e = 0$, they are given by
\beqa
\label{OmegaQprimestatic}
\Omega^0_{e^\prime} & = & - \vec Q_e\cdot \vec u_{e^\prime}\,,\nonumber\\
Q^0_{e^\prime} & = &  \sqrt{(\vec u_{e^\prime}\cdot \vec Q_e)^2 + Q^2_e}\,.
\eeqa
Therefore, there is a separate dependence on $\vec u_{e^\prime}\cdot\vec Q_e$,
and not just on the magnitude $Q_e$, and as a consequence the process
of taking the zero momentum limit $Q_e \rightarrow 0$ is not unique.

For our purposes (calculating the self-energy
in the presence of an external field), we
take the zero momentum limit in this case in the following manner.
First, after setting $\Omega_e = 0$, make an expansion of the stream
contribution to the vertex function in powers of
$\vec u_{e^\prime}\cdot\vec Q_e$, and then in harmony with
the treatment of the terms with $q_\mu$ specified above in \Eq{qasderivative},
interpret each such factor as a derivative
\beq
\label{derivativerule}
\vec u_{e^\prime}\cdot\vec Q_e \rightarrow
\frac{1}{i}\vec u_{e^\prime}\cdot\vec\nabla\,,
\eeq
acting on the external field. Since the remaining factors then depend
only on $Q_e$ after making this replacement, the $Q_e \rightarrow 0$
limit can be taken subsequently in an unambiguous way. In particular,
the stream contribution form factors, which are functions
of $\Omega_{e^\prime}$ and $Q_{e^\prime}$, are evaluated in this limit
according to a rule analogous to \Eq{staticlimit}, that is
\beq
\label{staticlimitprime}
(\Omega_{e^\prime} = 0, Q_{e^\prime} \rightarrow 0)\,.
\eeq
In this work we retain the terms that are at most
linear in the derivatives acting on the external magnetic
field after making the identifications made in
\Eqs{qasderivative}{derivativerule}.
In the idealized situation that the external field is
strictly homogeneous all such terms vanish.


For easy reference we quote here the following formulas that are given
in Eqs.\ (2.27-2.29) and Eq.\ (3.2) of ref.~\cite{DOlivo:1989ued},
\beqa
\label{ABCstatic}
A_f(0,Q_{f} \rightarrow 0) & = & A^0_f + O(Q^2_{f})\,,\nonumber\\
B_f(0,Q_{f} \rightarrow 0) & = & A^0_f + O(Q^2_f)\,,\nonumber\\
C_f(0,Q_{f} \rightarrow 0) & = & C^0_f + O(Q^2_f)\,,
\eeqa
where
\beqa
\label{AC0}
A^0_f & = & \frac{1}{2}\int\frac{d^3P}{(2\pi)^3}
\frac{\partial}{\partial\calE}\left[f_f(\calE) + f_{\bar f}(\calE)\right]\,,
\nonumber\\
C^0_f & = & \frac{1}{4}\int\frac{d^3P}{(2\pi)^3}\frac{1}{\calE}
\frac{\partial}{\partial\calE}\left[f_f(\calE) - f_{\bar f}(\calE)\right]\,.
\eeqa
In particular, $A_f(0,Q_f)$ and $B_f(0,Q_f)$ are equal at $Q_f = 0$, which
implies that $A^\prime_f(0,Q_f)$ and $T_{fT}(0,Q_f)$ are zero at $Q_f = 0$.
The derivation of the above formulas is sketched in
Appendix\ \ref{sec:ABCevalstaticlimit}, and
in Appendix \ref{sec:Aprimeevalstaticlimit} we derive
the formula for the static limit value of $A^\prime_f$,
including the $O(Q^2_f)$ term,
\beq
\label{Aprimestatic}
A^{\prime}_f(0, Q_f\rightarrow 0) = Q^2_f A^{\prime\,0}_f + O(Q^4_f)\,,
\eeq
where
\beq
\label{Aprime0}
A^{\prime\,0}_f = -\frac{1}{6}
\int\frac{d^3P}{(2\pi)^3}\frac{1}{\calE}
\frac{\partial}{\partial\calE}
\left[\frac{f_f(\calE) + f_{\bar f}(\calE)}{\calE}\right]\,,
\eeq
which will be relevant in the discussion in \Section{sec:selfenergy}.

The integrals defined in \Eqs{AC0}{Aprime0} can be performed straightforwardly
once the distribution functions are specified. For guidance and
reference purposes we give below the results of their evaluation
in the particular case that the distribution functions can be taken
to be those of the classical ideal gas. Using the fact that
in that the classical distribution function satisfies
\beq
\frac{\partial f}{\partial\calE} = -\beta f\,,
\eeq
(independently of whether the gas is relativistic or not), it follows
simply that
\beq
\label{A0classical}
A^0_f = -\frac{\beta_f}{4}(n_f + n_{\bar f})\,.
\eeq
In the case of $C^0$ and $A^{\prime\,0}$ the results
in the relativistic and non-relativistic cases
are different. In the non-relativistic limit ($\beta_f m \gg 1$)
\beqa
\label{CAprime0classicalNR}
C^0_f & = & -\frac{\beta_f}{8m}\left(n_f - n_{\bar f}\right)\,,\nonumber\\
A^{\prime\,0} & = & \frac{\beta_f}{12m^2}\left(n_f + n_{\bar f}\right)\,,
\eeqa
while in the extremely relativistic limit ($\beta_f m \ll 1$)
\beqa
\label{CAprime0classicalER}
C^0_f & = & -\left(\frac{\beta_f}{4}\right)^2 \left(n_f - n_{\bar f}\right)\,,
\nonumber\\
A^{\prime\,0}_f & = & \frac{\beta^3_f}{48}\sqrt{\frac{2\pi}{\beta_f m}}
(n_f + n_{\bar f})\,.
\eeqa
The derivation of \Eqs{CAprime0classicalNR}{CAprime0classicalER}
is sketched in Appendix\ \ref{sec:CAprime0classical}.

\section{Neutrino self-energy in a static magnetic field}
\label{sec:selfenergy}

\subsection{General form}
\label{subsec:generalform}

The chirality of the neutrino interactions implies that
the background contribution to the neutrino self-energy,
$\Sigma_{\mbox{eff}}$, is of the form
\beq
\Sigma_{\mbox{eff}} = R\Sigma L\,,
\eeq
and we will decompose $\Sigma$ in the form
\beq
\label{SigmamplussigmaB}
\Sigma = \Sigma^{(m)} + \Sigma^{(B)}\,,
\eeq
where $\Sigma^{(B)}$ stands for part that depends on $B$
and $\Sigma^{(m)}$ for the $B$-independent part.
The neutrino dispersion relation is obtained by solving the equation
\beq
\label{efffieldeq}
\left(\lslash{k} - \Sigma\right)\psi_L = 0\,.
\eeq

As is well known, in the two-stream electron background $\Sigma^{(m)}$
is of the form
\beq
\label{Sigmam}
\Sigma^{(m)} = \sum_f a_f\lslash{k} + \sum_f b_f \lslash{u}_f\,,
\eeq
where, to order $1/m^2_W$,
\beq
b_f = \frac{g^2}{4m^2_W}(n_f - n_{\bar f})\times\left\{
\begin{array}{ll}
1 + X_e & (\nu_e)\\
X_e & (\nu_{\mu,\tau})
\end{array}\right.
\eeq
The coefficients $a_f$ are $O(1/m^4_W)$ and therefore
we will discard them.

The issue that we address now is the enumeration of the possible
terms that can appear in $\Sigma^{(B)}$ for the two-stream background
with the external electromagnetic field. The situation we consider is that
in the rest frame of the normal background
there is a magnetic field $\vec B = B\hat b$, and in that frame we define
\beq
B^\mu = B b^\mu, \qquad b^\mu = (0, \hat b)\,.
\eeq
We can then write the corresponding EM tensor in the form
\beq
F_{\mu\nu} = \epsilon_{\mu\nu\alpha\beta} u^\alpha B^\beta \,,
\eeq
and its dual, as usual, is given by
\beqa
\label{Ftilde}
\tilde F_{\mu\nu} & = & \frac{1}{2}\epsilon_{\mu\nu\alpha\beta}F^{\alpha\beta}
\nonumber\\
& = & B_\mu u_\nu - u_\mu B_\nu\,.
\eeqa

In the enumeration of the possible terms that may appear in the
result of the 1-loop calculation of $\Sigma^{(B)}$,
we must remember the following working conditions:
\begin{enumerate}
\item restrict ourselves to the terms that are most linear
  in the derivatives of the field;
\item omit the terms that depend on the neutrino momentum $k$
  since they would be of $O(1/m^4_W)$ that we are not considering;
\item in the 1-loop calculation each background contributes independently,
  so that terms involving the products of vectors $u^\mu_f$ corresponding
  to different backgrounds do not appear.
\end{enumerate}
The following is then the list of the terms that can appear:
\begin{itemize}
\item[(a)] Terms with no derivatives of the electromagnetic field:
  $F^{\mu\nu} u_{f\nu}\gamma_\mu$
\item[(b)] Terms with one derivative of the electromagnetic field:
  \bdm
  \partial_\nu F^{\mu\nu}\gamma_\mu\;,
  (u_{f\alpha}\partial_\beta F^{\alpha\beta}) \lslash{u}_f\;,
  (u_f\cdot\partial F^{\mu\nu})u_{f\nu}\gamma_\mu
  \edm
\item[(c)] Terms similar to those in (a) and (b),
  with $F_{\mu\nu}\rightarrow\tilde F_{\mu\nu}$
\end{itemize}
Thus, under these conditions the most general form of $\Sigma^{(B)}$ is
\beqa
\label{SigmaBgeneral}
\Sigma^{(B)} & = & \sum_f\left[
  c_f \tilde F^{\mu\nu} u_{f\nu} +
  d_f F^{\mu\nu} u_{f\nu} +
  h_{f1} \partial_\nu F^{\mu\nu} +
  \tilde h_{f1} \partial_\nu \tilde F^{\mu\nu}\right.\nonumber\\
  &&\mbox{} +
  h_{f2} (u_{f\alpha}\partial_\beta F^{\alpha\beta}) u^\mu_f +
  \tilde h_{f2} (u_{f\alpha} \partial_\beta \tilde F^{\alpha\beta}) u^\mu_f
  \nonumber\\
  &&\mbox{} +
  \left.h_{f3}(u_f\cdot\partial F^{\mu\nu})u_{f\nu} +
  \tilde h_{f3}(u_f\cdot\partial \tilde F^{\mu\nu})u_{f\nu}\right]\gamma_\mu\,.
\eeqa
The coefficients defined here will be determined by calculating the
$\nu\nu$ transition amplitude in the presence of an external electromagnetic
field, using the 1-loop formulas for the vertex function given in
\Section{sec:vertexfunction}.

\subsection{$\nu\nu$ transition amplitude in an external electromagnetic field}
\label{subsec:nunutransitionamplitude}

\begin{figure}
{\centering
\resizebox*{0.35\textwidth}{0.2\textheight}
{\includegraphics{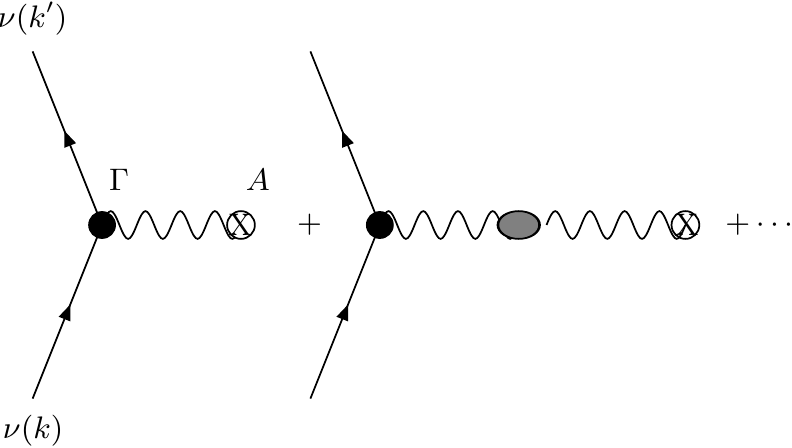}}
\par}
\caption{
Schematic diagrams for the effective neutrino
  electromagnetic vertex taking into account the polarization effects
  of the background electrons as expressed in \Eq{Snunu}.
  \label{fig2}
}
\end{figure}

We are now set to consider the $\nu\nu$ transition amplitude
in the presence of an external electromagnetic field. The
external electromagnetic potential is represented by
\beq
A_\mu(x) = \int\frac{d^4q^\prime}{(2\pi)^4}a_\mu(q^\prime)
e^{-iq^\prime\cdot x} \,,
\eeq
and the corresponding field by
\beq
F_{\mu\nu}(x) = \int\frac{d^4q^\prime}{(2\pi)^4}f_{\mu\nu}(q^\prime)
e^{-iq^\prime\cdot x} \,,
\eeq
where
\beq
\label{fmunu}
f_{\mu\nu}(q^\prime) = -i(q^\prime_\mu a_\nu(q^\prime) -
q^\prime_\nu a_\mu(q^\prime))\,.
\eeq
The diagram for the process is shown in  figure \ref{fig2}. As shown schematically
in that figure, it includes the photon polarization tensor in order to
take into account the screening effects of the background electrons.
%
%
%
The off-shell $\nu-\nu$ scattering amplitude in the presence the
external electromagnetic potential is then given by
\beq
\label{Snunu}
S_{\nu\nu} = -i\Gamma_{\lambda}(k,k^\prime) D^{\lambda\mu}(k^\prime - k)
a_\mu(k^\prime - k)\,,
\eeq
where $\Gamma_{\mu}$ is the total neutrino electromagnetic vertex function
given in \Eq{defGammaWZ} and
omitting the indices, the diagrams in figure \ref{fig2} indicate that
  $D = (1 + \Delta_0\pi + ...) = \Delta_{eff}\Delta^{-1}_0$, where
  $\Delta^{\mu\nu}_0 = \frac{-g^{\mu\nu}}{q^2}$ is the free photon
  propagator and $\Delta^{\mu\nu}_{eff}$ 
  is the effective photon propagator in the medium determined from
  \Eq{Deltaeffeq}.
%
%
%
\beq
\label{Dmunu}
D^{\mu\nu}(q) = -q^2 \Delta^{\mu\nu}_{eff}(q)\,,
\eeq
with $\Delta^{\mu\nu}_{eff}(q)$ being the photon propagator in the medium.
The latter quantity is determined by solving
\beq
\label{Deltaeffeq}
\left(q^2 \tilde g^{\mu\lambda} - \pi^{\mu\lambda}\right)
\Delta_{eff\,\lambda\nu} = -\tilde g^\mu_\nu\,,
\eeq
%
%
where $\pi_{\mu\nu}$ is the two-background contribution
to the photon polarization tensor. Denoting
by $\pi_{e\mu\nu}$ and $\pi_{e^\prime\mu\nu}$ the contributions
due to the normal and stream backgrounds, respectively, then
\beq
\label{pitotal}
\pi_{\mu\nu} = \pi_{e\mu\nu} + \pi_{e^\prime\mu\nu}\,.
\eeq
Each term in the previous relation can be decomposed according to
\beq
\pi_{f\mu\nu} = \pi_{fT} R_{f\mu\nu} + \pi_{fL} Q_{f\mu\nu}\,,
\eeq
where the photon self-energy functions $\pi_{fT,L}$ are given by
\beqa
\label{piLT}
\pi_{fT} & = & -2e^2\left[A_f - \frac{B_f}{\tilde u^2_f}\right]\nonumber\\
& = & -2e^2\left[A_f + \frac{q^2}{Q^2_f} B_f\right]\,,\nonumber\\
\pi_{fL} & = & -4e^2\frac{B_f}{\tilde u^2_f}\,,\nonumber\\
& = & 4e^2\frac{q^2}{Q^2_f} B_f\,,
\eeqa
with $A_{f}, B_{f}$ being the integrals defined in
\Eq{ABC}
(Omitting the subscript $f = e,e^\prime$, the longitudinal and transverse
  components of the dielectric constants in each background medium is
  are given by
  $\epsilon_{\ell} = 1 - \frac{\pi_{L}}{q^2}$ and
  $\epsilon_{t} = 1 - \frac{\pi_{T}}{\Omega^2}$
  respectively)

Let us look at the case considered in ref.~\cite{DOlivo:1989ued}, an external magnetic field
and only the normal background. In this case
\beq
\label{Deltaeffeqnormal}
\Delta^{\mu\nu}_{eff} = \frac{-R^{\mu\nu}_e}{q^2 - \pi_{eT}} +
\frac{-Q^{\mu\nu}_e}{q^2 - \pi_{eL}}\,,
\eeq
and
\beq
\label{Qaeq0}
Q_{e\mu\nu} a^\mu(q) = 0\,.
\eeq
\Eq{Qaeq0} follows from the fact that for a pure magnetic field
$\tilde u\cdot a = 0$, which can be seen in various ways.
For example, with the usual convention in which
$A^\mu = (0,\vec A)$ with $\nabla\cdot\vec A = 0$,
it follows that $u\cdot a = 0$ and $q\cdot a = 0$, and therefore
$\tilde u\cdot a = 0$. More generally,
$\tilde u\cdot a = \frac{i}{q^2} q_\mu u_\nu f^{\mu\nu}$, while
$u_\nu f^{\mu\nu} = 0$ when $f_{\mu\nu}$ corresponds to a magnetic field.
Thus, remembering that $\pi_{eT}\rightarrow 0$ (and $T_{eT}\rightarrow 0$)
in the static limit, \Eqs{Deltaeffeqnormal}{Qaeq0} imply that
$D^{\lambda\mu}a_\mu \rightarrow a^\lambda$ in the static limit,
and therefore the screening corrections are not relevant in that case.

With the stream contributions the situation is different.
The stream electrons, in their own rest frame, ``see''
an electric field in addition to the magnetic field, and the
screening corrections are relevant in that case.
The present situation is complicated by the fact that
in the presence of the two backgrounds the
inversion required in \Eq{Deltaeffeq} is not as simple in the general case
as the one leading to the one-background result given in \Eq{Deltaeffeqnormal}.

We overcome this difficulty here as follows. In the limit
$\vec u_{e^\prime} \rightarrow 0$,
each of the tensors $R_{e^\prime},Q_{e^\prime},P_{e^\prime}$ coincides
with its corresponding counterpart $R_{e},Q_{e},P_{e}$.
It is straightforward to show that, in general, the corresponding
primed and unprimed tensors differ by terms
$\sim (\vec u_{e^\prime}\cdot\vec Q_e)^2$, e.g.,
\beq
R_{e^\prime\mu\nu} = R_{e\mu\nu} +
O\left((\vec u_{e^\prime}\cdot\vec Q_e)^2\right)\,,
\eeq
with analogous relations for $Q_{e^\prime\mu\nu}$ and $P_{e^\prime\mu\nu}$.
Since, as stated in \Section{sec:introduction}, in this work we will retain
only up to the linear terms $\vec u_{e^\prime}\cdot\vec Q_e$ in the
calculation of the self-energy, we then can then write
\beqa
\label{screeningeq}
\Gamma_{\lambda}(k,k^\prime)\Delta^{\lambda}_{eff\mu}(q) &=&
T_{e\lambda\nu}\gamma^\nu L\left[
\frac{-R^\lambda_{e\mu}}{q^2 - \pi_T} +
\frac{-Q^\lambda_{e\mu}}{q^2 - \pi_L}\right]\nonumber\\
& +&
T_{e^\prime\lambda\nu}\gamma^\nu L\left[
\frac{-R^\lambda_{e^\prime\mu}}{q^2 - \pi_T} +
\frac{-Q^\lambda_{e^\prime\mu}}{q^2 - \pi_L}\right]\,,
\eeqa
where
\beqa
\pi_T & = & \pi_{eT} + \pi_{e^\prime T}\,,\nonumber\\
\pi_L & = & \pi_{eL} + \pi_{e^\prime L}\,.
\eeqa
We reiterate that \Eq{screeningeq} is valid assuming that we are dropping
the terms proportional to $(\vec u_{e^\prime}\cdot\vec Q_e)^2$ and higher
powers, which translate to terms containing second and higher derivatives
of the external field, by making the identification shown
in \Eq{derivativerule}. Using \Eq{defTTLP} and the multiplication
rules of the tensors $R, Q, P$, then
\beq
\label{GammaD}
\Gamma_{\lambda}(k,k^\prime) {D^{\lambda}}_{\mu}(q) =
\Gamma^{(eff)}_{e\mu}(k,k^\prime) +
\Gamma^{(eff)}_{e^\prime\mu}(k,k^\prime)\,,
\eeq
where
\beqa
\Gamma^{(eff)}_{f\mu}(k,k^\prime) & = & \left[
\tilde T_{fT}R_{f\mu\nu} +
\tilde T_{fL}Q_{f\mu\nu} -
\tilde T_{fP}P_{f\mu\nu}
\right]\gamma^\nu L \,,
\eeqa
with
\beqa
\tilde T_{fT,P} & = & \frac{q^2 T_{fT,P}}
{q^2 - \pi_{T}}\,,\nonumber\\
\tilde T_{fL} & = & \frac{q^2 T_{fL}}
{q^2 - \pi_{L}}\,.
\eeqa
%
%
%
Using \Eqs{Snunu}{GammaD}, the $\nu\nu$ amplitude is then given by
\beq
\label{Snunu2}
S_{\nu\nu} = -i\left(\Gamma^{(eff)}_{e\mu}(k,k^\prime) +
\Gamma^{(eff)}_{e^\prime\mu}(k,k^\prime)\right) a^\mu(k^\prime - k)\,.
\eeq
An equivalent expression for the functions $\Gamma^{(eff)}_{f\mu}$,
which is more convenient for the purpose of the interpretation
of the form factors and for taking the static limit, is \cite{DOlivo:1989ued}
\beqa
\label{formfactorparam}
&&\Gamma^{(eff)}_{f\mu}(k,k^\prime) =
\left[F_{f1}\tilde g_{\mu\nu}\gamma^\nu +
F_{f2}\tilde u_{f\mu}\lslash{u}_f \right.\nonumber\\
&&\left .+
iF_{f3}(\gamma_\mu u_{f\nu} - \gamma_\nu u_{f\mu})q^\nu +
iF_{f4}\epsilon_{\mu\nu\alpha\beta}
\gamma^\nu q^\alpha u^\beta_f\right]L\,,
\eeqa
where, using the formulas given in \Eq{defRQP} for the tensors
$R_{f\mu\nu}, Q_{f\mu\nu}, P_{f\mu\nu}$, 
\beqa
\label{FTTLPequivalence}
F_{f1} & = & \tilde T_{fT} +
\frac{\Omega^2_f}{Q^2_f}(\tilde T_{fL} - \tilde T_{fT})\,,\nonumber\\
F_{f2} & = & \frac{1}{\tilde u^2_f}
(\tilde T_{fL} - \tilde T_{fT})\,,\nonumber\\
iF_{f3} & = & -\frac{\Omega_f}{Q^2_f}
(\tilde T_{fL} - \tilde T_{fT})\,,\nonumber\\
F_{f4} & = & \frac{\tilde T_{fP}}{Q_f} \,.
\eeqa
%
%
%
%
It follows from \Eq{formfactorparam} that
\beq
\label{GammaMXf}
\Gamma^{(eff)}_{f\mu}(k,k^\prime) a^\mu(k^\prime - k) =
M_{f\mu\nu}f^{\mu\nu}(k^\prime - k)\,,
\eeq
where $f^{\mu\nu}$ is defined in \Eq{fmunu} and the $M_{f\mu\nu}$
are given by
%
%
%
\beqa
\label{defMXf}
M_{f\mu\nu} &=& \left[
-i\frac{F_{f1}}{q^2} q_\mu \gamma_\nu -
i\frac{F_{f2}}{q^2} q_\mu u_{f\nu}\lslash{u}_f\right.\nonumber\\
&&\left.  +
F_{f3} u_{f\mu}\gamma_\nu  -
\frac{1}{2}F_{f4}\epsilon_{\mu\nu\alpha\beta}u^\alpha_f \gamma^\beta 
\right]L\,,
\eeqa
and from \Eq{Snunu2}
\beq
\label{Snunu3}
S_{\nu\nu} = -i
\left(M_{e\mu\nu} + M_{e^\prime\mu\nu}\right) f^{\mu\nu}(k^\prime - k)\,.
\eeq
\Eq{Snunu3} is a useful starting point to determine
the contribution to the neutrino self-energy in a static field,
including the case of an inhomogeneous field, which we consider next.

\subsection{Self-energy}
  
We now consider the $\nu\nu$ transition amplitude
for the case of a static magnetic field $F_{\mu\nu}$.
As stated in \Section{subsec:staticlimit}, we make
the idealization that we are calculating it over a region that
is microscopically large but macroscopically sufficiently
small such that the external field and its space derivatives are
constant over that region. In addition we retain only the terms
that are most linear in the derivatives. Operationally this means
that in \Eq{Snunu3} we can take
\beqa
\label{staticfieldrules}
f_{\mu\nu}(k^\prime - k) & = & (2\pi)^4 \delta^{(4)}(k^\prime - k) F_{\mu\nu}
\,,\nonumber\\
q_\lambda f_{\mu\nu}(k^\prime - k)
& = & (2\pi)^4 \delta^{(4)}(k^\prime - k) i\partial_\lambda F_{\mu\nu}\,,
\eeqa
while neglecting the terms with higher powers of $q$, which would translate
to terms with higher order derivatives of the external field.

To state our working assumptions more precisely let us
denote by $\Delta x$ the distance over which the magnetic
field $B$ changes appreciably. Since the variation of B over a given distance
$\delta x$ is
$
\delta B = \left(\frac{\partial B}{\partial x}\right)\delta x\,,
$
and $\Delta x$ is determined by the condition that
$
\delta B \sim B\,,
$
we have
\begin{displaymath}
\Delta x \sim \frac{B}{\left(\frac{\partial B}{\partial x}\right)}\,.
\end{displaymath}
In the calculations, as in every QFT calculation,
we idealize a region (of linear size $L$) that is microscopically
large ($L \gg \lambda$) compared to the neutrino Compton wavelength
$\lambda = \frac{1}{k}$,
such that it is valid to take the usual $L \rightarrow \infty$ limit
(or $\lambda/L \rightarrow 0$). If $\lambda$ is sufficiently small such that
$
\lambda \ll L \ll \Delta x
$
can be satisfied, then the field $B$ can be taken as constant
over the region and in such cases the first formula given
in \Eq{staticfieldrules} is strictly valid, which is the usual
homogeneous field case.
In the present paper we assume that $L$ is not necessarily much smaller
than $\Delta x$,
in which case we cannot take the field as being constant over the region
$L$. What we assume by adopting the formulas in \Eq{staticfieldrules} is
that the field variations can be treated perturbatively, so that we
can keep only the leading term in a Taylor series expansion in each
formula. In cases in which the inhomogeneities of the background
medium are important on a level comparable to the homogeneous
background, this assumption would not hold and \Eq{staticfieldrules}
is not valid. Our calculations and treatment hold under the assumption
that the inhomogeneities are small and whence can be taken into account
by a perturbative treatment, in the sense just indicated.

Under the conditions we have stated, \Eq{Snunu3} then becomes 
\beq
\label{Snunuext}
S_{\nu\nu} = -i (2\pi)^4 \delta^{(4)}(k - k^\prime)\left(
M^{(static)}_{e\mu\nu} + M^{(static)}_{e^\prime\mu\nu}\right)F^{\mu\nu}\,,
\eeq
where $M^{(static)}_{f\mu\nu}$ is obtained from \Eq{defMXf} by
keeping the terms that are at most linear in
$(\vec u_{e^\prime}\cdot \vec Q_e)$ and/or $q$ and following
the procedure outlined in \Section{subsec:staticlimit}: make the
identification stated in \Eqs{qasderivative}{derivativerule},
and then take the $q\rightarrow 0$ limit
in the remaining terms as indicated in \Eqs{staticlimit}{staticlimitprime}.
The $B$-dependent part of
the self-energy, $\Sigma^{(B)}$, which is identified by writing
\beq
S_{\nu\nu} = -i (2\pi)^4 \delta^{(4)}(k - k^\prime)\Sigma^{(B)}L\,,
\eeq
is then given by
\beq
\Sigma^{(B)} = \Sigma^{(B)}_e + \Sigma^{(B)}_{e^\prime}\,,
\eeq
where
\beq
\Sigma^{(B)}_{f} = M^{(static)}_{f\mu\nu}F^{\mu\nu}\,.
\eeq
Calculating $M^{(static)}_{f\mu\nu}$ as we have indicated,
%
%
%
\beqa
\label{SigmaBfinal}
\Sigma^{(B)}_{f} &=& 
\left[t_{fT}\partial_\nu F^{\mu\nu} +
t_{fL} (u_{f\alpha} \partial_\beta  F^{\alpha\beta})u^\mu_{f} \right. \nonumber\\
&&\left.
-
t_{fL}(u_f\cdot\partial\, F^{\mu\nu}) u_{f\nu}  +
t_{fP} \tilde F^{\mu\nu}u_{f\nu}\right]\gamma_\mu L \,,
\eeqa
where the coefficients $t_{fT,L,P}$ are defined by
\beqa
\label{tTLPdef}
t_{fT} & = &  \left.\frac{T_{fT}(0,Q)}{Q^2}\right|_{Q\rightarrow 0}
\,,\nonumber\\[12pt]
t_{fL} & = & \left.\frac{T_{fL}(0,Q)}
{\pi_{eL}(0,Q) + \pi_{e^\prime L}(0,Q)}\right|_{Q\rightarrow 0}
\,,\nonumber\\[12pt]
t_{fP} & = & \left.\frac{T_{fP}(0,Q)}{Q}\right|_{Q\rightarrow 0}\,.
\eeqa
Using \Eqsss{TTLPformulas}{ABCstatic}{Aprimestatic}{piLT}
we obtain the following explicit formulas,
\beqa
\label{tTLP}
t_{fT} & = & \frac{eg^2}{2m^2_W}A^{\prime\,0}_f \times \left\{
\begin{array}{ll}
  1 + X_e & (\mbox{$\nu_e$})\\[12pt]
  X_e & (\mbox{$\nu_{\mu,\tau}$})
\end{array}\right.\nonumber\\[12pt]
t_{fL} & = & \frac{-g^2}{4e m^2_W}\left(
\frac{A^0_f}{A^0_e + A^0_{e^\prime}}
\right)
\times \left\{
\begin{array}{ll}
  1 + X_e & (\mbox{$\nu_e$})\\[12pt]
  X_e & (\mbox{$\nu_{\mu,\tau}$})
\end{array}\right.\nonumber\\[12pt]
t_{fP} & = & \frac{eg^2}{m^2_W}C^0_f \times \left\{
\begin{array}{ll}
  1 - Y_e & (\mbox{$\nu_e$})\\[12pt]
  -Y_e & (\mbox{$\nu_{\mu,\tau}$})
\end{array}\right.
\eeqa
where $A^0_f, C^0_f, A^{\prime\,0}_f$ are the integrals defined in
\Eqs{AC0}{Aprime0}.

\Eqs{SigmaBfinal}{tTLP} summarize the result of our calculation of the
contribution to the neutrino self-energy due to its interaction with
an external electromagnetic field that is not necessarily homogeneous.
The result given in \Eq{SigmaBfinal} corresponds to the
general form given in \Eq{SigmaBgeneral}, with the
coefficients given specifically by
\beqa
\label{htrelations}
c_f & = & t_{fP}\,,\nonumber\\
h_{f1} & = & t_{fT}\,,\nonumber\\
h_{f2} & = & t_{fL}\,,\nonumber\\
h_{f3} & = & -t_{fL}\,,\nonumber\\
d_f = \tilde h_{f1} = \tilde h_{f2} = \tilde h_{f3} & = & 0\,.
\eeqa
%

\section{Dispersion relations}
\label{sec:dispersionrelation}

For the purpose of determining the dispersion relation, it is
convenient to express the total self-energy, \Eq{SigmamplussigmaB},
in the form
\beq
\Sigma = \lslash{V}\,,
\eeq
with
\beq
V^\mu  = \sum_f V^\mu_f\,.
\eeq
The formula for the $V^\mu_f$ follows from \Eqs{Sigmam}{SigmaBgeneral},
which we summarize in the form
\beq
\label{defVf}
V^\mu_f = V^{(h)\mu}_f + V^{(i)\mu}_f\,,
\eeq
where
\beqa
\label{defVhif}
V^{(h)\mu}_f & = & b_f u^\mu_f + c_f \tilde F^{\mu\nu} u_{f\nu}\,,\nonumber\\
V^{(i)\mu}_f & = & h_{f1} \partial_\nu F^{\mu\nu} +
h_{f2} (u_{f\alpha}\partial_\beta F^{\alpha\beta}) u^\mu_f \nonumber\\
&&+
h_{f3}(u_f\cdot\partial F^{\mu\nu})u_{f\nu}\,.
\eeqa
$V^{(i)\mu}_f$ is non-zero only if the field is inhomogeneous.
In writing \Eq{defVhif} we have dropped the terms that vanish
according to the results we have obtained in \Eq{htrelations}.
We can express $V^{(i,h)\mu}_f$ more explicitly as,
\beqa
\label{Vihfexplicit}
V^{(h)}_{f\mu} & = & b_f u_{f\mu} +
c_f\left[(u_f\cdot u)B_\mu - (u_{f}\cdot B) u_\mu\right]\,,\nonumber\\
V^{(i)\mu}_f & = & -h_{f1}m^\mu - h_{f2}(u_f\cdot m)u^\mu_f
\nonumber\\
&&-
h_{f3}\epsilon^{\mu\nu\alpha\beta} u_{f\nu} u_\alpha n_{f\beta}\,.
\eeqa
In the expression for $V^{(h)}_{f\mu}$ we have used \Eq{Ftilde},
and for $V^{(i)}_{f\mu}$ we have introduced the vectors
\beqa
m^\mu & = & \partial_\lambda F^{\lambda\mu}\,,\nonumber\\
n^\mu_{f} & = & -(u_f\cdot\partial)B^\mu\,, 
\eeqa
which in the rest frame of the matter background have components
\beqa
m^\mu & = & (0, \vec m) \,,\nonumber\\
n^{\mu}_f & = & (0, \vec n_f) \,,
\eeqa
where
\beqa
\vec m & = & \nabla\times\vec B\,,\nonumber\\
\vec n_f & = & (\vec u_f\cdot\nabla)\vec B\,.
\eeqa
The equation for the propagating modes, \Eq{efffieldeq},
implies that the dispersion relations are given by
\beq
k^0 - V^0 = \pm\left|\vec k - \vec V\right|\,.
\eeq
Thus to the lowest order in $1/m^2_W$ considered in this work,
which among other things implies that $V^\mu$ does not depend on $k$,
and the solutions are $k^0 = \omega_{\pm}(\vec k)$, where
\beq
\label{omegapm}
\omega_{\pm}(\vec k) = \pm\left[|\vec k| - \hat k\cdot\vec V\right] + V^0\,.
\eeq
In \Eq{omegapm} $\hat k$ denotes the unit vector along the direction of
propagation.
The neutrino and antineutrino dispersion relations, which are
identified in the usual way as,
\beqa
\omega_\nu & = & \omega_{+}(\vec k)\,,\nonumber\\
\omega_{\bar\nu} & = & -\omega_{-}(-\vec k)\,,
\eeqa
are then given by
\beq
\label{omeganubarnu}
\omega_{\nu,\bar\nu} = |\vec k| \pm \delta\,,
\eeq
where
\beq
\label{delta}
\delta = \sum_f(V^0_f - \hat k\cdot\vec V_f) \,.
\eeq
According to the decomposition in \Eq{defVf}, we can write
\beq
\label{deltafdecomp}
\delta = \sum_f(\delta^{(h)}_f + \delta^{(i)}_f)\,,
\eeq
with
\beq
\delta^{(h,i)}_f = V^{(h,i)0}_f - \hat k\cdot\vec V^{(h,i)}_f \,,
\eeq
and from \Eq{Vihfexplicit},
\beqa
\label{deltahif}
\delta^{(h)}_f & = & b_f u^0_f + c_f\vec B\cdot \vec u_f
- b_f \hat k\cdot\vec u_f
- c_f u^0_f \hat k\cdot B\,,\nonumber\\
\delta^{(i)}_f & = & h_{f1}(\hat k\cdot\vec m) +
h_{f2} u^0_f (\vec u_f\cdot\vec m) -
h_{f2}(\vec u_f\cdot\vec m)(\hat k\cdot\vec u_f) \nonumber\\
&&+
h_{f3}\hat k\cdot(\vec u_f\times \vec n_f)\,.
\eeqa
%

%
%
%
For the two-stream electron background specifically,
using \Eqsto{defu}{defv} and \Eq{htrelations},
\beqa
\label{deltahtwostream}
\delta^{(h)}_e & = & b_e - c_e\hat k\cdot\vec B\,,\nonumber\\
\delta^{(h)}_{e^\prime} & = & 
b_{e^\prime} v^0 + c_{e^\prime}\vec v\cdot\vec B - 
b_{e^\prime} \hat k\cdot\vec v - c_{e^\prime} v^0\hat k\cdot\vec B\,,
\eeqa
and
\beqa
\label{deltaitwostream}
\delta^{(i)}_e & = & t_{eT}(\hat k\cdot\vec m)\,,\nonumber\\
\delta^{(i)}_{e^\prime} & = & t_{e^\prime L} v^0 (\vec v\cdot\vec m) +
t_{e^\prime T}(\hat k\cdot\vec m) -
t_{e^\prime L}(\vec v\cdot\vec m)(\hat k\cdot\vec v)\nonumber\\
&&-
t_{e^\prime L}\hat k\cdot(\vec v\times \vec n_{e^\prime})\,,
\eeqa
%
%
%
where
\beq
\label{defneprime}
\vec n_{e^\prime} = (\vec v\cdot\nabla)\vec B\,.
\eeq
In this case,
\beq
\label{deltatwostream}
\delta = \delta^{(h)}_e + \delta^{(h)}_{e^\prime} + \delta^{(i)}_e+ \delta^{(i)}_{e^\prime}\,.
\eeq
%

 
Together with \Eq{omeganubarnu}, \Eqs{deltafdecomp}{deltahif}
provide a general and concise expression for the neutrino
and antineutrino dispersion relations
in the kind of situation that we are envisaging,
and \Eqss{deltahtwostream}{deltaitwostream}{deltatwostream} in particular 
for the two-stream electron background we are specifically considering.
In these formulas the stream velocity $\vec v$ is left unspecified
since we do not consider the possible physical origin of the
stream background. However, the results can be used in specific
applications in which the stream velocity is determined and/or
restricted by the particular physical conditions of the problem. For example,
if the stream  velocity is due to the drift of electrons in the B field,
since the Lorentz force makes charged particles drift only along the
$\vec B$ axis but not in the perpendicular plane, the
results can be applied to that case as well by taking $\vec v$ to be on the
$\vec B$ axis. But as we have just stated,
we do not assume anything in particular about the origin of the streams
or what determines their velocities, and the results hole for more general
cases as well in which the stream velocity need not be along the
magnetic field. 

It is useful to consider some special situations
that illustrate particular features of the general results.

\subsection{Homogeneous external field}

It is simple to verify that the results obtained in \cite{Nieves:2017rvx}
are reproduced as a special case. Thus, if the external field is homogeneous,
\beq
\label{deltatwostreamconstB}
\delta = \delta^{(h)}_e + \delta^{(h)}_{e^\prime}\,,
\eeq
which is the result obtained in \cite{Nieves:2017rvx}.
In particular, in the absence of the stream background,
\beq
\label{deltae}
\delta = b_e - c_e \hat k\cdot\vec B\,,
\eeq
which leads by \Eq{omeganubarnu} to the result obtained in ref.~\cite{DOlivo:1989ued}
for the dispersion relation in a magnetized electron background.
The angular asymmetry of the dispersion relation in this case
has been subject of much interest in the literature in connection to
the problem of pulsar kick and related issues.
The terms in \Eq{deltahtwostream} due to the stream background
(in the same case of a homogeneous field) give an additional
asymmetric contribution that depends on the direction of propagation
relative to the stream background velocity.

\subsection{Inhomogeneous magnetic field}

\subsubsection{$\nabla\times\vec B = 0$}

In the case of a non-homogeneous field, the additional terms
given by $\delta^{(i)}_{e^\prime}$ can be present.
As an example, let us consider the case in which $\nabla\times\vec B = 0$.
In this case, only the term involving $h_{e^\prime 3}$ in \Eq{deltaitwostream}
survives and therefore, from \Eq{deltatwostream},
\beqa
\delta &=& b_e + c_{e^\prime}\vec B\cdot\vec v + b_{e^\prime} v^0 -
b_{e^\prime} \hat k\cdot\vec v \nonumber\\
&&- \left(c_e + c_{e^\prime} v^0\right)
\hat k\cdot\vec B -
t_{e^\prime L}\hat k\cdot(\vec v\times\vec n_{e^\prime})\,,
\eeqa
where $\vec n_{e^\prime}$ has been defined in \Eq{defneprime}.
In particular, in addition to the angular dependence involving
the direction of propagation relative to $\vec B$ and $\vec v$, there
is a dependence involving a third vector $\vec v\times\vec n_{e^\prime}$.

\subsubsection{$\nabla\times\vec B \not= 0$}

On the other hand, if the conditions are such that $t_{e^\prime L}$
terms in \Eq{deltatwostream} are negligible, then
\beqa
\delta &=& b_e + c_{e^\prime}\vec B\cdot\vec v + b_{e^\prime} v^0 -
b_{e^\prime} \hat k\cdot\vec v \nonumber\\
&&- \left(c_e + c_{e^\prime} v^0\right)
\hat k\cdot\vec B + (t_{eT}+t_{e^\prime T}) (\hat k\cdot\vec m)\,.
\eeqa
Notice in particular that even in the absence of the stream, in which case
\beq
\delta = b_e - c_e \hat k\cdot\vec B + t_{eT}(\hat k\cdot\vec m)\,,
\eeq
there is an additional anisotropic term of the form
$\hat k\cdot (\nabla\times\vec B)$, besides the usual one proportional
to $\hat k\cdot\vec B$.


\subsection{Discussion}

One point that stands out in these results is the following.
To the order $1/m^2_W$, the coefficients $b_{e,e^\prime}, c_{e,e^\prime}$
are proportional to the electron-positron asymmetry in the corresponding
backgrounds. Therefore in a $CP$-symmetric medium $b_{e,e^\prime}$ and
$c_{e,e^\prime}$ vanish to the order $1/m^2_W$, and
in such cases the order $1/m^4_W$ contributions to these parameters must
be included. In contrast, to the order $1/m^2_W$, the $t_{eT}$
term in $\delta^{(i)}_e$, and similarly the $t_{e^\prime L,T}$ terms
in $\delta^{(i)}_{e^\prime}$,
are proportional to the sum of the electron-positron number
densities (in the normal and stream backgrounds respectively),
and whence need not be zero even in a $CP$-symmetric medium.
Thus, in a $CP$-symmetric medium (e.g., the Early Universe)
the dominant contribution to the neutrino index of refraction
could be due to the terms contained in $\delta^{(i)}_{e,e^\prime}$,
which are of order $1/m^2_W$.


To quantify somewhat these statements, recall that the
$O(1/m^4_W)$ contribution to $b_e$ is \cite{DOlivo:1992lwg}
\beq
\label{b4}
b^{(4)}_e \sim \frac{g^2 |\vec k|T_e}{m^4_W} N_e\,,
\eeq
where
\beq
N_e = n_e + n_{\bar e}\,,
\eeq
and similarly for $b^{(4)}_{e^\prime}$.
\Eq{b4} holds for the case the electron gas can be adequately
described by a classical distribution in the relativistic limit ($T_e \gg m$).
For illustrative purposes the question we ask is under what conditions the
term $t_{e^\prime L}\hat k\cdot(\vec v\times {\bf n_{e^\prime}})$ could be as
important, or more important, than the $b^{(4)}_e$ term in the dispersion
relation. Under the example idealized conditions that we have assumed for the
purpose of this discussion (classical relativistic electron backgrounds),
using \Eqs{A0classical}{tTLP} the question translates to see
for what parameters it is possible that
\beq
\frac{g^2}{em^2_W}
\left(\frac{\beta_{e^\prime} N_{e^\prime}}{\beta_{e} N_{e} +
  \beta_{e^\prime} N_{e^\prime}}\right)|\nabla \vec B|
\sim \frac{g^2 E_\nu T_e}{m^4_W}N_e \,,
\eeq
where we have put $E_\nu \sim |\vec k|$.
Taking the quantity in parenthesis to be $O(1)$ and $N_e \sim T^3_e$,
the condition would be that
\beq
\frac{1}{e}|\nabla\vec B| \sim \frac{T^4_e E_\nu}{m^2_W} \,,
\eeq
which yields
\beq
\frac{1}{e}|\nabla\vec B| \sim
10^3 \left(\frac{T_e}{MeV}\right)^4 \left(\frac{E_\nu}{MeV}\right)
\frac{MeV^2}{meter}\,,
\eeq
or
\beq
|\nabla\vec B| \sim \left(\frac{T_e}{MeV}\right)^4
\left(\frac{E_\nu}{MeV}\right)\frac{10^{15}G}{cm}\,.
\eeq

We can ask the same question for the term $t_{e^\prime T}(\hat k\cdot\vec m)$.
Using \Eqs{CAprime0classicalER}{tTLP}, the condition that this term
be of the same order as $b^{(4)}_e$ would be
\beq
\frac{eg^2}{4m^2_W}\frac{\beta^3_{e^\prime}}{48}
\sqrt{\frac{2\pi}{\beta_{e^\prime} m}}N_{e^\prime}|\nabla\vec B| \sim
\frac{g^2 E_\nu T_e}{m^4_W}N_e \,,
\eeq
or, taking again $N_f \sim T^3_f$,
\beq
|\nabla\vec B| \sim \left(\frac{T_e}{MeV}\right)^4
\left(\frac{T_{e^\prime}}{MeV}\right)^{-1/2}
\left(\frac{E_\nu}{MeV}\right)\frac{10^{17} G}{cm}\,.
\eeq
Similarly, the condition for the term $t_{eT}(\hat k\cdot\vec m)$
to be comparable to $b^{(4)}_e$ is given by this same relation,
with the replacement $T_{e^\prime}\rightarrow T_e$.

On the other hand, if the medium is electron-positron asymmetric, the effects
of the inhomogeneous terms seem to be unimportant compared to the
standard terms. As an specific example, let us compare the
$t_{e^\prime T}(\hat k\cdot\vec m)$ against the term $c_e B$, which is also
a source of an asymmetry in the dispersion relation.
The condition that it be of the same order as $c_e B$ is
\beq
\frac{eg^2}{4m^2_W}\frac{\beta^3_{e^\prime}}{48}
\sqrt{\frac{2\pi}{\beta_{e^\prime} m}}N_{e^\prime}|\nabla\vec B| \sim
\frac{eg^2}{2m^2_W}\left(\frac{\beta_e}{4}\right)^2\Delta N_e B\,,
\eeq
or
\beqa
\label{cecompare}
\frac{|\nabla\vec B|}{B} & \sim & \left(\frac{T_{e^\prime}}{T_e}\right)^2
\sqrt{m T_{e^\prime}}\left(\frac{\Delta N_{e}}{N_{e^\prime}}\right)\nonumber\\
& \sim & \left(\frac{T_{e^\prime}}{T_e}\right)^2
\sqrt{\frac{T_{e^\prime}}{MeV}}\left(\frac{\Delta N_{e}}{N_{e^\prime}}\right)
10^{11} cm^{-1}\,,
\eeqa
where we have used \Eqs{tTLP}{htrelations}, and defined
$\Delta N_e = n_e - n_{\bar e}$. 
Similarly, comparing $t_{eT}(\hat k\cdot\vec m)$ against $c_e B$
would require
\beq
\frac{|\nabla\vec B|}{B}
\sim \sqrt{\frac{T_{e}}{MeV}}\left(\frac{\Delta N_{e}}{N_{e}}\right)
10^{11} cm^{-1}\,.
\eeq
Thus if we assume, for example, that $\Delta N_{e} \sim N_{e}
\sim N_{e^\prime}$ and $T_e \sim T_{e^\prime}$, the conditions become
%
%
%
\beq
\frac{|\nabla\vec B|}{B} \sim
\sqrt{\frac{T_{e}}{MeV}}\; 10^{11} cm^{-1}\,.
\eeq
As these particular cases illustrate, the contributions
to the dispersion relation due to the inhomogeneity of the magnetic field
do not seem to be significant if the background is electron-positron
asymmetric.

%
%
%
%

However, more generally, it is not inconceivable that
those contributions may be relevant under the appropriate environmental
conditions including an electron-positron symmetric background.
While we have not considered a specific application, the example estimates
above show that the gradient-dependent contributions could be comparable
to the standard terms in such environments. Since they give rise to
distinctive kinematic features in the dispersion relation
(e.g., angular asymmetries) the possible need to include their
effects in some specific application contexts should be kept in mind.  

\section{Conclusions and Outlook}
\label{sec:conclusions}

In this work we have calculated the electromagnetic vertex function
of a neutrino that propagates in a medium consisting 
of a \emph{normal} electron background plus another
electron \emph{stream} background which is moving with a 
velocity relative to the normal background.
The results obtained were used to determine the neutrino self-energy
and dispersion relation in such a medium in the presence of
an external magnetic field ($\vec B$), paying special attention
to the case in which $B$ is inhomogeneous, keeping the terms
that are linear in $B$ and its spatial derivatives.

The 1-loop formulas for the vertex function were given in
\Section{sec:vertexfunction}. The calculation is based on
the local limit of the weak interactions,
that is, it is restricted to the order $1/m^2_W$ terms only.
The formulas generalize those given in \cite{DOlivo:1989ued},
adapted to the present context.
The vertex function is expressed in terms of a set of
form factors that are given as integrals over the distribution functions
of the background electrons. We also summarized the static limit
value of the integrals involved in such formulas, which are subsequently
used in the calculation of the neutrino self-energy and dispersion relation
in the two-stream medium in the presence of a non-homogeneous external field.

In \Section{sec:selfenergy} we used the results for the vertex function
in the two-stream medium to determine neutrino self-energy
in the presence of a static external magnetic field. In contrast
to the previous calculations of the neutrino index of refraction in
magnetized media, we took into account and emphasized the case
in which the field is not homogeneous. There we explained in
some detail the need to include the screening effects of the
background electrons in the calculation of the self-energy
in the two-stream medium case.
The results for the $B$-dependent part of the self-energy
are summarized in \Eqs{SigmaBfinal}{tTLP}.

The corresponding dispersion relations were obtained and discussed
in \Section{sec:dispersionrelation}, focusing on some of the features
that depend on the inhomogeneity of the $B$ field
and/or the presence of the stream electron background.
In the presence of an
inhomogeneous field the dispersion relation acquires
additional anisotropic terms, in particular one of the form
$\hat k\cdot(\nabla\times \vec B)$, which is independent of the stream
background velocity and can be present even in the absence of the stream
background, and other terms, such as the gradient of
$\hat k\cdot(\vec v\times\vec B)$, that
depend on the stream background velocity and can be
present even in the case in which $\nabla\times \vec B = 0$.
As we showed, the terms that depend on the field derivatives,
in contrast to those that depend on $B$ itself,
are proportional to the sum of the electron and positron densities,
and therefore are non-zero to order $1/m^2_W$ in a $CP$-symmetric
medium in which the particle and antiparticle densities are equal.
Thus, in a $CP$-symmetric medium
the dominant contribution to the neutrino index of refraction
could be due to the terms that depend on the derivatives of $\vec B$,
which are of order $1/m^2_W$, in contrast with the constant $B$ terms
which to that order vanish and are of order $1/m^4_W$ in that case.

From a more general point of view, the present work is a step
in our effort to study problems related to the propagation
of neutrinos in a medium that consists of various particle backgrounds
that may be streaming with different velocities. The results of our first step
in this direction were presented in \cite{Nieves:2017rvx}, in which
we considered the propagation of a neutrino
in a magnetized two-stream electron background medium. There
we considered the self-energy and dispersion
relation of neutrino that propagates in a two-stream
electron medium, that is a medium composed
of an electron background taken to be at rest (to which we refer as the normal
background), and a second electron background that moves with some velocity 
relative to the first. In addition we assumed that, in the rest frame
of the normal background, there is a magnetic field that is homogeneous.
Here we have extended that work by considering the neutrino
electromagnetic vertex function in the two-stream electron medium.
As already mentioned in the Introduction,
the knowledge of the vertex function allows
us to determine the neutrino electromagnetic properties and to
calculate the rate for various processes involving
neutrinos in such media, but we do not consider these applications here.
Alternatively, by considering the effective
neutrino interaction with an external electromagnetic field, we have
used the results for the vertex function to determine the self-energy
and dispersion relation of a neutrino that propagates in a two-stream medium
with an inhomogeneous magnetic field. In particular this extends the previous
works on neutrino propagation in magnetized media which are restricted
to the case of a homogeneous magnetic field.
There is an extensive literature related to the effects of an external
magnetic field in the propagation of neutrinos in dense media
in a variety of astrophysical and cosmological contexts.
The results of this work provide a firm setting for exploring
the effects of the combined presence of stream backgrounds and
inhomogeneities of external fields along the same lines, which can be
applicable in real astrophysical and cosmological situations, such as:
pulsars, supernova, gamma-ray bursts and Early Universe as already 
mentioned in the introduction.


We are thankful to the anonymous referee for his insightful comments.
S.S is thankful to Japan Society for the promotion of science (JSPS)
for the invitational fellowship. The work of S.S. is partially supported
by DGAPA-UNAM (M\'exico) Project No. IN110815 and PASPA-DGAPA, UNAM.

\appendix
\section{Evaluation of the integrals $A_f, B_f, C_f$ in the static limit}
\label{sec:ABCevalstaticlimit}

The integrals to be evaluated are those given in \Eq{ABC}.
In the static limit [\Eqs{staticlimit}{staticlimitprime}],
the functions $A_f, B_f, C_f$ are then given by
\beqa
A_f(0,Q_{f}) & = & -\int\frac{d^3P}{(2\pi)^3 2\calE}
(f_{f}(\calE) + f_{\bar f}(\calE))\nonumber\\
&&\times
\left[\frac{2m^2 + 2\vec P\cdot\vec Q_{f}}{Q^2_f +
2\vec P\cdot\vec Q_{f}} +
(\vec Q_{f}\rightarrow -\vec Q_{f})\right]
\,,\nonumber\\
B_f(0,Q_{f}) & = & -\int\frac{d^3P}{(2\pi)^3 2\calE}
(f_{f}(\calE) + f_{\bar f}(\calE)) \nonumber\\
&&\times
\left[\frac{2\calE^2 + \vec P\cdot\vec Q_{f}}
{Q^2_{f} + 2\vec P\cdot\vec Q_{f}} +
(\vec Q_{f}\rightarrow -\vec Q_{f})\right]
\,,\nonumber\\
C_f(0,Q_{f}) & = & -\frac{1}{2}\int\frac{d^3P}{(2\pi)^3}
(f_{f}(\calE) - f_{\bar f}(\calE)) \nonumber\\
&&\times
\left[\frac{1}{Q^2_{f} + 2\vec P\cdot\vec Q_{f}} +
(\vec Q_{f}\rightarrow -\vec Q_{f})\right]\,,
\eeqa
where
\beqa
\calE & = & p\cdot u_f\,,\nonumber\\
P & = & \sqrt{\calE^2 - m^2} \,.
\eeqa
Both the normal and stream background terms are evaluated in the static
limit in similar fashion. As discussed in \Section{sec:introduction}
and summarized in \Eqs{staticlimit}{staticlimitprime}, in both cases
we must first set $\Omega_f = 0$ and subsequently take the limit
$Q_f\rightarrow 0$, in that order.

For illustrative purposes let us consider $A_f(0,Q_{f})$ first
in some detail. Making the change of variable
\beq
\label{changevarP}
\vec P \rightarrow \vec P \mp \frac{1}{2}\vec Q_{f}
\eeq
in the first and second terms inside the square bracket, respectively,
\beqa
\label{Astatic1}
A_f(0,Q_{f}) &=& -\int\frac{d^3P}{(2\pi)^3}
\frac{1}{2\vec P\cdot\vec Q_{f}}
\left[\left(m^2 - \frac{1}{2}Q^2_{f}\right)\left(N_{-} - N_{+}\right)\right.
\nonumber\\
&& \left.+
\vec P\cdot\vec Q_{f}\left(N_{-} + N_{+}\right)\right]
\eeqa
where we have defined
\beq
\label{NP}
N(\vec P) \equiv \frac{(f_f(\calE) + f_{\bar f}(\calE))}{\calE} \,,
\eeq
and
\beq
\label{Nplusminus}
N_{\pm} = N\left(\vec P \pm \frac{1}{2}\vec Q_{f}\right)\,.
\eeq
Since we are interested in eventually taking the limit
$\vec Q_{f}\rightarrow 0$, we expand
\beqa
\label{Napprox}
N\left(\vec P \pm \frac{1}{2}\vec Q_{f}\right) & = &
N \pm \frac{1}{2}\vec Q_{f}\cdot\nabla_P N + O(Q^2_f)\nonumber\\
& = & N \pm \frac{1}{2}\vec Q_{f}\cdot\vec P
\frac{1}{\calE}\frac{\partial N}{\partial\calE} + O(Q^2_{f})\,,
\eeqa
where in the second line we have used the fact that $N$
depends on $\calE$ (and not explicitly on $\vec P$) so that
\beq
\label{gradpidentity}
\nabla_P N  = \frac{\vec P}{\calE}\frac{\partial N}{\partial\calE}\,.
\eeq
Substituting \Eq{Napprox} in \Eq{Astatic1} we obtain
\beq
A_f(0,Q_{f}) = -\frac{1}{2}\int\frac{d^3P}{(2\pi)^3}\left\{
-m^2\frac{1}{\calE}\frac{\partial N}{\partial\calE} + 2N\right\} + O(Q^2_{f})\,.
\eeq
Writing $m^2 = \calE^2 - \vec P^{\,2}$ and using \Eq{gradpidentity}
once more, the term in curly brackets above can be rewritten,
\beq
-m^2\frac{1}{\calE}\frac{\partial N}{\partial\calE} + 2N =
-\frac{\partial(\calE N)}{\partial\calE} + \nabla_P\cdot(\vec P N)\,,
\eeq
and since the second term on the right-hand side integrates to zero we thus
finally obtain the result quoted in \Eq{ABCstatic}.
The results for $B_f$ and $C_f$ quoted in \Eq{ABCstatic}
can be obtained in similar fashion.
In particular, since $A_f$ and $B_f$ are equal at $Q_f = 0$, it follows
that $A^\prime_f$ is zero at $Q_f = 0$, which in turns implies that
$T_{fT}(0,Q_f) = O(Q^2_f)$. In Appendix \ref{sec:Aprimeevalstaticlimit}
we calculate the $O(Q^2_f)$ term of $A^\prime_f$.
%

\section{Evaluation of the integral $A^\prime_f$ in the static limit}
\label{sec:Aprimeevalstaticlimit}

Here we calculate the static limit value of $A^\prime_f$ including
the $O(Q^2_f)$ term, which in turn determines $T_{fT}(0,Q_f)$ and in the end
$t_{fT}(0,Q_f)$ as defined in \Eq{tTLP}. We wish to state explicitly here
that the momentum thermal distribution functions are assumed
to be isotropic (in each background's rest frame), so that the symmetric
integration replacements such as that in \Eq{symmint} below are valid.
Thus the result given in \Eq{Aprimefinal} [which is quoted in
\Eq{Aprimestatic}] is subject to this restriction. From \Eq{ABC},
\beq
A^\prime_f(0, Q_f) = -\frac{1}{2}\int\frac{d^3P}{(2\pi)^3}N(\vec P)
\left[\frac{-2\vec P^2 + \vec P\cdot\vec Q_f}{Q^2_f + 2\vec P\cdot\vec Q_{f}} + 
(\vec Q_{f}\rightarrow -\vec Q_{f})\right]\,,
\eeq
and making the change of variable indicated in \Eq{changevarP},
\beqa
\label{Aprimestatic1}
A^\prime_f(0, Q_f) & = & -\frac{1}{2}\int\frac{d^3P}{(2\pi)^3}
\frac{1}{2\vec P\cdot\vec Q_f}\left[3\vec P\cdot\vec Q_f(N_{-} +
  N_{+})\right.
\nonumber\\
&&\left.
-(2 P^2 + Q^2_f)(N_{-} - N_{+})\right]\nonumber\\
& = & -\frac{1}{2}\int\frac{d^3P}{(2\pi)^3}\; 3N - \frac{1}{2}
\tilde A^\prime\,,
\eeqa
where we have defined
\beq
\label{Aprimetilde}
\tilde A^\prime = 
-\int\frac{d^3P}{(2\pi)^3}
\frac{1}{2\vec P\cdot\vec Q_f}\left[(2 P^2 + Q^2_f)(N_{-} - N_{+})\right]\,,
\eeq
and we have used the definitions given in \Eqs{NP}{Nplusminus} once more.
In this case we expand $N_{\mp}$ up to the cubic terms in $\vec Q_f$,
\beq
\label{cubicexpansionN}
N_{-} - N_{+} = -\vec Q_f\cdot\nabla_P N - \frac{1}{3}
\left(\frac{Q^i_{f}Q^j_{f}Q^k_{f}}{8}\right)\nabla^i_P \nabla^j_P \nabla^k_P N
+ O(Q^5_f)\,,
\eeq
and obtain
\beqa
\tilde A^\prime & = & \int\frac{d^3P}{(2\pi)^3}
\frac{1}{2\vec P\cdot\vec Q_f}(2P^2 + Q^2_f)\left[
\vec Q_f\cdot\nabla N + \frac{1}{3\cdot 8}\left(\vec Q_f\cdot\nabla\right)^3
N\right]\nonumber\\
& \equiv & I^{(0)} + I^{(2)}\,,
\eeqa
where
\beqa
\label{AprimeI0I2}
I^{(0)} & = & \int\frac{d^3P}{(2\pi)^3}
\frac{\vec P^2}{\vec P\cdot\vec Q_f}(\vec Q_f\cdot\nabla N)
\,,
\nonumber\\
I^{(2)} & = & \int\frac{d^3P}{(2\pi)^3}\frac{1}{2\vec P\cdot\vec Q_f}
\left[Q^2_f (\vec Q_f\cdot\nabla N) \right.\nonumber\\
&&\left. +
\frac{2}{3\cdot 8} P^2 (\vec Q_f\cdot\nabla)^3 N\right]\,.
\eeqa
Using \Eq{gradpidentity},
\beq
I^{(0)} = \int\frac{d^3P}{(2\pi)^3}\;\vec P\cdot\nabla N\,,
\eeq
and therefore, from \Eq{Aprimestatic1},
\beq
\label{AprimeintermsofI2}
A^\prime(0, Q_f) = -\frac{1}{2}I^{(2)}\,,
\eeq
where have used the fact that
\beq
3N + \vec P\cdot\nabla N = \nabla\cdot(\vec PN)\,,
\eeq
which integrates to zero and therefore does not contribute to $A^\prime$.

For the evaluation of $I^{(2)}$, we first
write all the derivatives with respect to $P^i$ in terms of derivatives
with respect to $\calE$ using \Eq{gradpidentity}, which we express in the form
\beqa
\label{nablaiN}
\nabla^i_P N & = & P^i N^\prime\,,\nonumber\\
\nabla^i_P \nabla^j_P N & = & \delta^{ij} N^\prime +
P^i P^j N^{\prime\prime}\,,\nonumber\\
\nabla^i_P \nabla^j_P \nabla^k_P N & = & (P^i\delta^{jk} + P^j\delta^{ik} +
P^k\delta^{ij}) N^{\prime\prime} \nonumber\\
&&
+ P^i P^j P^k N^{\prime\prime\prime}\,,
\eeqa
where we define
\beqa
N^\prime & = & \frac{1}{\calE}\frac{\partial N}{\partial\calE}\,,\nonumber\\
N^{\prime\prime} & = &
\frac{1}{\calE}\frac{\partial}{\partial\calE}
\left(\frac{1}{\calE}\frac{\partial N}{\partial\calE}\right)\,,
\eeqa
and so on.
Using \Eq{nablaiN}, the first term in the integral $I^{(2)}$ in \Eq{AprimeI0I2}
is reduced using
\beq
\label{I2firstterm}
\frac{1}{\vec P\cdot\vec Q_f}\vec Q_f\cdot\nabla N = N^\prime\,,
\eeq
while for the second term we use
\beqa
\label{Qcubereduction}
\frac{1}{\vec P\cdot\vec Q_f}(\vec Q_f\cdot\nabla)^3 N & = &
3 Q^2_f N^{\prime\prime} + (\vec P\cdot\vec Q_f)^2 N^{\prime\prime\prime}
\nonumber\\
& \rightarrow & Q^2_f\left[
3 N^{\prime\prime} + \frac{1}{3}P^2 N^{\prime\prime\prime}\right]\,.
\eeqa
The second line in \Eq{Qcubereduction} indicates the replacement
that can be made in the integral over $\vec P$, which allows us to replace
\beq
\label{symmint}
P^i P^j \rightarrow \frac{1}{3}P^2\delta^{ij}\,,
\eeq
in the integrand. Thus $I^{(2)}$ becomes
\beq
I^{(2)} = Q^2_f\int\frac{d^3P}{(2\pi)^3}\left\{\frac{1}{2} N^\prime +
\frac{1}{3\cdot 8}P^2
\left[3 N^{\prime\prime} + \frac{1}{3}P^2 N^{\prime\prime\prime}\right]
\right\}\,.
\eeq
This integral can be simplified further by using the following identities
\beqa
\label{Nprimerelations}
P^2 N^{\prime\prime} & \rightarrow & -3 N^\prime\,,\nonumber\\
P^2 P^2 N^{\prime\prime\prime} & \rightarrow & -5 P^2 N^{\prime\prime}
\nonumber\\
& \rightarrow & 5\cdot 3 N^\prime\,,
\eeqa
where the arrow symbol means that the relations hold under the integral
sign after dropping the terms that integrate to zero. \Eq{Nprimerelations}
can be shown by contracting the second formula in \Eq{nablaiN}
with $\delta^{ij}$ and the third one with $P^i\delta{jk}$,
\beqa
\nabla^2 N & = & 3N^\prime + P^2 N^{\prime\prime}\,,\nonumber\\
(\vec P\cdot\nabla) \nabla^2 N & = & 5 P^2 N^{\prime\prime} +
P^2 P^2 N^{\prime\prime\prime}\,,
\eeqa
and noticing that the left-hand side in each of them
can be written as a total divergence and therefore integrate to zero.
Therefore, using the same notation,
\beq
P^2\left[3 N^{\prime\prime} + \frac{1}{3}P^2 N^{\prime\prime\prime}\right]
\rightarrow -4 N^\prime\,,
\eeq
so that,
\beqa
I^{(2)} & = & \left[\frac{1}{2} - \frac{4}{3\cdot 8}\right]Q^2 N^\prime
\nonumber\\
& = & \frac{1}{3}Q^2N^\prime\,,
\eeqa
and from \Eq{AprimeintermsofI2} we arrive at the final result
quoted in \Eq{Aprimestatic},
\beq
\label{Aprimefinal}
A^\prime_f(0,\vec Q_f) = -\frac{Q^2_f}{6}
\int\frac{d^3P}{(2\pi)^3}\frac{1}{\calE}
\frac{\partial}{\partial\calE}
\left[\frac{f_f(\calE) + f_{\bar f}(\calE)}{\calE}\right] + O(Q^4_f)\,.
\eeq

\section{Evaluation of $A^{\prime\,0}_f, C^0_f$ in the classical limit}
\label{sec:CAprime0classical}

Let us consider $A^{\prime\,0}$, defined in \Eq{Aprime0}.
Performing the
angular integration, and using
\beq
\frac{\partial}{\partial\calE} = \frac{\calE}{P}\frac{\partial}{\partial P}\,,
\eeq
followed by an integration by parts,
\beq
\label{A0primeclass1}
A^{\prime\,0} = \frac{1}{12\pi^2}\int dP\frac{1}{\calE}
(f_f(\calE) + f_{\bar f}(\calE))\,.
\eeq
Let us consider the integral over $f_f(\calE)$, where
\beq
f_{f}(\calE) = e^{\alpha_f}e^{-\beta_f \calE}\,,
\eeq
in the classical limit that we are considering. Making the change of variable
\beq
P = m\cosh\xi\,,
\eeq
\beqa
\label{A0primeclass2}
\int dP\frac{1}{\calE} f_f(\calE) & = & e^\alpha_f\int d\xi
e^{-\beta_f m\cosh\xi}\nonumber\\[12pt]
& \simeq & \sqrt{\frac{\pi}{2\beta m}}e^\alpha_f e^{-\beta m}\nonumber\\[12pt]
& = & \left\{
\begin{array}{ll}
  \frac{\pi^2\beta_f}{m^2} n_f & (\beta_f m \gg 1)\\[12pt]
  \frac{\pi^2\beta^3_f}{4}\sqrt{\frac{2\pi}{\beta_f m}} n_f &
  (\beta_f m \ll 1)
\end{array}
\right.
\eeqa
The second step follows by expanding $\cosh\xi$ up to the quadratic term
in $\xi$ and performing the Gaussian integration, and in the third step
we have used the explicit relationship between the total number density $n_f$
and the chemical potential $(e^\alpha)$, which is different in the
non-relativistic ($\beta_f m \gg 1$) and the extremely-relativistic
($\beta_f m \ll 1$) limits. The integral over $f_{\bar f}$ can be expressed
in analogous fashion.

Thus, using the result given in \Eq{A0primeclass2} (and the corresponding
result for $f_{\bar f}$) in \Eq{A0primeclass1} we arrive
at the formulas for $A^{\prime\,0}$ quoted in
\Eqs{CAprime0classicalNR}{CAprime0classicalER}. The formulas for $C^0$
quoted in there follow in similar fashion.
%
%



%

\end{document}